\documentstyle[11pt,epsfig]{article}

\title{\bf Quantum gravity: a solution for current problems in cosmology and particle physics}
\author{F. Darabi\thanks{email: f.darabi@azaruniv.edu }
\\{\small  Department of Physics, Azarbaijan
University of Tarbiat Moallem, Tabriz 53741-161, Iran}\\{\small
Research Institute for Astronomy and Astrophysics of Maragha
(RIAAM), Maragha 55134-441, Iran}}
\begin{document}
\maketitle
\begin{abstract}
We propose a new phenomenological model for quantum gravity. This is
based on a new interpretation in which quantum gravity is not an
interaction, rather it is just responsible for generation of
space-time-matter. Then we show this model is capable of solving the
important problems of cosmology and particle physics.
\\
{\bf PACS: 04.60.Bc; 12.10.Kt ; 98.80.Cq; 98.80.Es}\\
{\bf Keywords: quantum gravity; Planck scales; discrete space-time;
mass hierarchy; cosmological constant; early universe}\\
\end{abstract}
\vspace{2cm}
\newpage
\section{Introduction}

The ``problem of time'' in General Relativity is of particular
importance which means the theory is a completely time parametrized
system in the sense that there is no natural notion of time due to
the diffeomorphism invariance of the theory and therefore the
canonical Hamiltonian as the generator of time reparametrizations
vanishes which casts in the form of Hamiltonian constraint. Physical
observables are functions of phase space which are reparametrization
invariant, that is, they commute with all Dirac constraints and they
do not evolve with respect to the canonical Hamiltonian. One
important problem of time comes from cosmology and can be explained
as follows: The Friedmann equations describe the physical time
evolution which is actually observed for instance through red shift
experiments. The puzzle here is that these observed quantities as
functions of the phase space do not commute with the constraints, so
they are not gauge invariant and therefore should not be observable,
in sharp contradiction to observation. Moreover, the Friedmann
equations are generated by a vanishing Hamiltonian constraint and
must be interpreted as gauge transformations rather than evolution
equations. Therefore, the phase space dependent quantities in these
equations should not be observable, again in sharp contradiction to
reality. The conclusion is: Either the constraint formalism which
has been exactly tested by experiments in other gauge theories is
inappropriate or we are missing some new physics in gravitation
theory \cite{0}.

There is another important problem in the standard model called the
``Hierarchy problem''. It is known that at least two fundamental
energy scales exist in nature and they are extremely different: the
Planck scale $M_{P}\sim 10^{19}$ GeV and the electroweak scale
$M_{EW}\sim 10^3$ GeV, with the ratio around $M_{P}/M_{EW}\sim
10^{16}$. Over the past two or three decades there was a great
interest to explain this large ratio which is known as the Hierarchy
problem. One may translate this problem into the puzzle that why
gravity is so weak compared to the other interactions, such as
electroweak one. So, it seems the resolution of this problem will
lead us to a better understanding of our Universe. Many proposals to
solve this problem have been made so far. By his Large number
hypothesis, Dirac first conjectured that the Newton's constant may
be a time-dependent parameter $G\sim t^{-1}$ to solve the problem
\cite{1}. Recently, a great amount of interest has been concentrated
on solving this problem based on the existence of extra dimensions
in brane models. These proposals are based on the higher-dimensional
unification in that the gravitational force in higher-dimension is
supposed to be as strong as other forces, but a relatively large
internal space of the order of TeV scale or a "warp" factor which is
a rapidly changing function of an additional dimension makes the
4-dimensional gravitational force very weak \cite{2}. Some attempts
have also been done to solve this problem based on the
non-commutativity in the space-time coordinates or even in the
mini-superspace coordinates of specific quantum cosmological models
\cite{3}. Recently, an idea based on Mach principle has also been
introduced to solve this problem with the price that one has to
resort to weak form of equivalence principle and assume the inertial
and gravitational masses of any object are not exactly the same
\cite{5'}.

The third problem is known as the so called cosmological constant
problem. The problem arises in the following way
\begin{equation}
\Lambda_{eff} = \Lambda_0 + {O}(M_P)^{4} \label{0}
\end{equation}
where $\Lambda_{eff}$ is a combination of bare cosmological
constant $\Lambda_0$ and quantum field theory contributions of the
order of quantum gravity cut off $(M_P)^4$. Due to huge values of
these contributions we face with the so-called cosmological
constant problem, namely a large difference of the order
$10^{120}$ between the observational bound and theoretical
predictions on the value of cosmological constant.

To the authors opinion, such enigmatic problems which have resisted
a long time against being solved by the most ``advanced'' theories
and ``sophisticated'' technics imply that we may need to turn back
and look for ``simple'' and ``natural'' solutions to these problems
just in our ``observed'' four dimensional universe with
space-time-matter elements\footnote{We point that the idea in
\cite{5'} is a step forward to solve the hierarchy problem within
our four dimensional universe. However, the idea which is introduced
here is more fundamental and rich enough to solve the other problems
of particle physics and cosmology as well as hierarchy problem, in a
more fashionable form.}. For example, instead of a model based on
``complex'' higher dimensional picture of the universe to solve the
hierarchy and cosmological constant problems we may suspect about
some ``complex'' evaluations of different energy scales, without
considering their four dimensional ``length-time'' scales which may
lead to those unjustifiable large numbers. In this regard, it seems
a correct interpretation of quantum gravity in which length, time
and mass-energy scales are well defined may solve the important
current problems in cosmology and particle physics.

In this paper we propose a new idea on the phenomenology of quantum
gravity which is mainly based on the assumption that quantum gravity
is not a framework to quantize the gravitational interaction as
mediating a force, rather it is a mechanism by which quantum baby
universes with length-time-mass scales as the building blocks of the
whole universe are continuously generated. We will show how this
model may solve the current problems of cosmology and particle
physics.

\section{Discrete space-Time }

The notion of discrete space-time is not a new one. Some people have
already proposed such a discrete notion of time or space-time
\cite{4}. However, no physical mechanism by which space-time can be
discretized is ever introduced. It is important to note that any
such mechanism must respect gravity. In fact, since general
relativity as a gravitational theory very respects the important
role of space-time, the quantum gravity is the only framework which
is deeply involved with the space-time structure at small scales.
The time problem, whose solution is not given by general relativity,
is therefore strongly supposed to be solved within the context of
quantum gravity.

In any constraint system with some gauge invariant quantities one
may fix the gauge freedom. The important gauge freedom in general
relativity is space-time reparametrization. So, one may arbitrarily
fix the lapse and shift functions in the metric to fix the gauge.
However, this kind of gauge fixing does not help to solve the time
problem, in particular at the cosmological level, as explained
above. A very important step in this direction may be the assumption
that ``space-time'' is a {\it discrete} notion having a fundamental
unit which is mainly fixed by quantum gravity. In other words, we
will assume that quantum gravity is nothing but a phenomena by which
these fundamental units are generated. This self-gauge-fixing in a
consistent way makes all the gravitational quantities to be
naturally defined in terms of this fundamental unit of space-time (a
fundamental gauge). A fundamental unit of space-time requires a
fundamental parametrization in space-time, which is different from
what is usually done by arbitrarily fixing, for example the shift or
lapse functions.

There is only one space-time scale in nature which is uniquely fixed
by gravitation, namely Planck length-time
\begin{equation}
l_P=\sqrt{\frac{G \hbar}{c^3}}, \:\:\: t_P=\sqrt{\frac{G
\hbar}{c^5}},\label{1}
\end{equation}
where $G$ is the gravitational constant, $\hbar$ is the planck
constant and $c$ is the constant velocity of light. We will call
this set as our fundamental {\it quantum} unit of space-time which
discretizes the space-time continuum. Note that $G$ as an element of
gravity fixes the length and time scales in a unique and fundamental
way. At classical or even quantum mechanical level space-time is
just a continuum, but at scales close to Planck length-time scales,
it becomes fundamentally discrete and can not be arbitrarily
reparametrized. In fact, the reason why general relativity is a
reparametrization invariant theory is that space-time at the level
of general relativity is a continuum made of huge numbers of quantum
units of space-time and we can easily change the time parameter or
shift the space coordinates by arbitrarily small values due to the
additive properties of these quantum units. However, going down to
the Planck scales, which are the characteristic features of quantum
gravity regime, we can no longer reparametrize the space-time,
simply because a fundamental gauge fixing is made by ``quantum
gravity'' itself through the fundamental constants $G$, $\hbar$, and
$c$ in (\ref{1}) and one can not breaks down the fundamental quantum
unit of space-time. According to above discussion, we will assume
that the ``gravitational'' universe has its specific language
(gauge) based on Planck natural units, $t_P$, $l_P$, and $M_P$. In
other words, we will compare any time, length and mass scales in the
gravitational universe just with Planck time, length and mass.

\section{Quantum gravity, time problem and space-time-matter generation}

At time scales close to $t_P$ and length scales close to $l_P$
``quantum gravity'' is supposed to become important, just like
``space'' and ``time'' whose discrete natures become important at
these scales. Suppose, instead of thinking about quantum gravity as
a ``complicate framework'' of quantizing the background metric
$g_{\mu \nu}$ with a quantum particle so called {\it graviton}
mediating the gravitational force \cite{Wald}, we think about it as
a ``simple phenomena'' by which fundamental quantum units of
space-time with the Planck length-time scales, namely ``quantum
 baby universes'' are just generated. This is a very important step
forward leaving behind all the technical and conceptual problems
concerning the quantization of background metric $g_{\mu \nu}$
\footnote{An idea was already developed as the ``gravitational
complementary principle'' by the author \cite{5} to show how we can
absolutely separate the domains of quantum gravity from general
relativity, not only in their energy scales but also in their
essence. In fact, it was shown that general relativity is not the
low energy limit of quantum gravity, rather it is absolutely ``pure
classical'' gravitational theory which governs the scales larger
enough than Planck length and there is no absolutely a quantum
theory of gravity at large scales. For example, it was shown that
gravitational waves as the metric perturbations may exist in the
framework of general relativity but no (quantum) particle concept
like graviton can be attributed to this wave behavior. In other
words, no wave-particle duality is expected for gravitational
radiation. In the present model, we almost follow the same picture
and show that the pure classical features of gravity manifest as
general relativity at large scales, and that quantum gravity is just
responsible for continuous creation of the universe by Planck
length-time scales quantum baby universes. Moreover, no Heisenberg
uncertainty principle, namely wave-particle duality is expected for
these quantum baby universes in an expanding universe.}. In the
present model: we suppose a ``virtual'' quantum baby universe of
Planck length-time-mass is initially born, with no real
characteristics, by a vacuum quantum gravitational fluctuation
obeying uncertainty principle. Each single quantum baby universe
generated possibly by successive vacuum quantum gravitational
fluctuations is also supposed to be a {\it virtual} one. However, we
assume once they (at least two virtual quantum baby universes) can
correlate and interplay with each other by ``Mach principle'' they
all get realized into a larger ``mother'' universe with real
characteristics imposed by Mach principle. Therefore, by this
assumption we certify that: our present {\it real} large scale four
dimensional space-time whose origin is assumed to be a Planck scales
{\it virtual} universe ( created during a quantum gravitational
fluctuation in the vacuum state) is nothing but a huge reproduction
of such virtual baby universes which have been steadily realized
into the mother universe, according to Mach principle\footnote{We
may assume the spatial part of each quantum baby universe is a
spherical shell (bubble) with the tickness of one Planck length. The
internal radius of each bubble is always equal to the radius of the
universe. So, at early universe the size of each quantum baby
universe (bubble) is so smaller than that of at present universe. Of
course, the relative size of quantum baby universe is irrelevant in
our model, rather its absolute tickness does matter.}. Realization
into mother universe of such single virtual universes each of which
created by uncertainty principle from the vacuum, is done by Mach
principle. Therefore, Mach principle in the case of quantum gravity
plays a very important role to avoid these newborn (through
uncertainty principle) virtual quantum baby universes from
annihilation to the vacuum. If there was no Mach principle, the
uncertainty principle would annihilate each virtual quantum baby
universe after its creation and would not let a large scale mother
universe to be realized by huge accumulation of these virtual
quantum baby universes\footnote{This phenomena is like to the pair
production out of electromagnetic vacuum at the horizon of a black
hole. The horizon lets the newborn virtual pair particles to be
realized. If the horizon would not exist these virtual particles
would annihilate again to the electromagnetic vacuum, by uncertainty
principle.}. The interesting point is that once, for example, two
such virtual quantum baby universes get realized they share
uncertainty principle in a Machian way so that this Machian
universe, namely mother universe as a whole, inherit the uncertainty
principle. In other words, uncertainty principle is just applied for
the whole ``real Machian mother universe'' and not for each quantum
ingredient as `` virtual single baby universe'', because after
realization into the mother universe they have no longer a definite
and specific length-time-mass characteristics for
themselves\footnote{We know that general relativity as a fully
deterministic theory has a sharp conflict with uncertainty principle
so that one can determine simultaneously the position and velocity
of a particle locally in general relativity. If we assume that the
large scale universe is constructed by huge numbers of quantum baby
universes then locally it is like to a quantum baby universe. This
is based on the large scale-small scale duality, connecting the
large scale universe to the quantum baby universe, which is
established by Mach-Heisenberg principle. Therefore, in order the
locally deterministic behavior of general relativity is realized,
each virtual quantum baby universe as a local universe should behave
deterministically rather than obeying uncertainty principle,
otherwise one can not localize a particle in the deterministic way
that general relativity requires. In fact, due to the Machian
correlations (socialization) between the virtual quantum baby
universes (with single characters) we can no longer expect each
virtual single quantum baby universe to keep its original single
character and obey uncertainty principle after it is introduced into
this Machian mother universe as a realized community constructed by
accumulation of virtual single quantum baby universes. Therefore,
one can only expect the Machian universe ``as a whole'', no matter
what size and age, to obey the uncertainty principle. There is no
fundamental problem if we leave uncertainty principle for a single
local quantum baby universe, namely the Planck scale unit of
space-time, after its realization into the Machian universe. The
situation is almost the same when we leave special relativity for
the expansion of the universe. Expansion of the universe with
velocities more than velocity of light does not contradict the
special relativity to be applied to natural relative velocities. In
the same way, a single quantum baby universe (included in the mother
universe) which does not obey the uncertainty principle locally,
does not contradict this principle to be applied globally to the
mother universe, as a whole. The uncertainty principle applies just
for the whole ``real mother universe'' not for its ingredients as ``
virtual quantum baby universes''. This follows exactly because of
space-time expansion of the universe. The position-momentum and
time-energy uncertainty relations are usually defined on a fixed
space-time background. Therefore, it is reasonable to apply these
relations on the three gauge interactions (weak, strong,
electromagnetic) which are usually assumed to take place on a fixed
background. However, in the case of general relativity we are
dealing with expanding background (space-time), so we can no longer
expect these relations to be valid locally for a quantum unit of
space-time-matter. In an expanding universe, for example given by
(\ref{2}), the expansion of the space-time boundaries of the
universe renormalizes every $\Delta x$ and $\Delta t$ corresponding
to general relativity, by imposing decaying measures as
$\tilde{\Delta}x=10^{-60}\Delta x$ and
$\tilde{\Delta}t=10^{-60}\Delta t$ so that $\tilde{\Delta}x$ and
$\tilde{\Delta}t$ are decreasing by expansion of the universe. One
may then reduce simultaneously $\Delta p$-$\tilde{\Delta}x$ or
$\Delta E$-$\tilde{\Delta}t$ to arbitrarily small values, for each
quantum unit of space-time-matter. This certifies our conjecture
that uncertainty principle looses gradually its local validity for
each quantum baby universe because of the global expansion of the
universe. In other words, in the case of general relativity the
expansion of the universe kills the local validity of uncertainty
principle for each quantum baby universe while keeping it valid
globally for the mother universe. But, the global expansion of the
universe is due to the gradual accumulation of quantum baby
universes. It then appears that addition of one quantum baby
universe to the mother universe decreases globally the quantum
importance of other quantum baby universes already existed in the
mother universe and rises their classical importance. Therefore, as
the universe expands the quantum local features of space-time,
namely uncertainty principle or wave-particle duality, is lost in
favor of pure classical features \cite{5}. For example, the
``metric'' in a large enough universe becomes almost a pure
classical field with no probabilistic features. This solves the
contradiction between deterministic features of general relativity
and probabilistic features of uncertainty principle in favor of the
former. The mechanism by which the uncertainty principle looses its
local validity for a quantum baby universe in an expanding universe
while it is being valid globally for the mother universe is very
much like the formation of a community. In a one-person community
the normalization factor (social value) of this person is one. In a
two-persons community the normalization factor reduces to one-half.
In a $10^{60}$-persons community the normalization factor reduces to
$10^{-60}$ while the system of $10^{60}$-persons (mother universe)
is still a ``comm-unity'' (uni-verse) having the same original
one-person character (uncertainty principle). It is important to
note that the realization of each new quantum baby universe is a
global effect throughout the mother universe because this
realization is ``recognized'' by the global Mach-Heisenberg
principle which is imposed on the mother universe. Therefore, for
example, the reduction of cosmological constant due to the
realization of new quantum baby universes is a global effect (see
below).} Therefore, if for example the energy scale of the original
quantum baby universe is of the Planck mass $M_P$, the energy scale
of the enlarged Machian mother universe should be reduced below the
Planck mass due to Heisenberg uncertainty principle which is now
applied for this enlarged mother universe. An important result is
that if in a vacuum quantum gravitational fluctuation during which a
unit of space-time of Planck length-time orders is realized into the
mother universe, a corresponding energy scale could also be realized
into this mother universe, its order of magnitude should be at most
of the same energy scale of the mother universe, namely reduced
Planck energy. In other words, once this enlarged universe is
realized through Mach-Heisenberg principle, its energy scale is
reduced below the Planck energy and just an energy scale at most of
the order of reduced Planck energy can be borrowed from vacuum and
then be realized as a real energy or mass into this mother Machian
universe. This procedure may continue non stoping to enlarge more
and more the Machian mother universe and reduce less and less its
energy scale. So, this enlarging universe at each stage of its time
evolution can borrow a decreased energy, corresponding to the energy
scale of that stage, from vacuum to convert it to a real energy or
mass. Due to the existence of fundamental quantum gravitational
natural units of length-time-energy, namely the Planck scales it is
very natural to expect that the enlargement of universe in
space-time and the subsequent decay of its energy scale, is realized
in terms of Planck units.

We assume that the enlargement or ``time evolution of the universe''
is nothing but ``steadily realization of virtual quantum universes
of Planck length-time generated by vacuum quantum gravitational
fluctuations. In specific words, when ``each'' vacuum quantum
gravitational fluctuation generates ``one'' virtual unit of
space-time which is then realized as a quantum universe of Planck
length-time scales, we say that the ``age'' of the mother universe
gets older up to one Planck time and its ``radius'' gets larger up
to one Planck length. Therefore, although at the level of general
relativity we have an apparent continuous time reparametrization
symmetry with no natural notion of time, but at ultrashort distances
this symmetry is broken by the discrete structure of space-time and
so a preferred time parameter is singled out as a natural one $t=N
t_P$ where $N$ is an integer indicating the numbers of quantum
universes or fundamental units of space-time generated dynamically
so far by vacuum quantum gravitational fluctuations in the entire
history of the universe. In this way, time evolution of the universe
becomes a dynamical process controlled by vacuum quantum
gravitational fluctuations so that cosmological quantities when
interpreted in terms of these quantum fluctuations become
meaningful. This solves the ``problem of time'' at the cosmological
level by resorting to such a quantum gravity model.

We showed, according to the Mach-Heisenberg principles, that as the
original quantum baby universe becomes large and larger, its energy
scale in contact with the gravitational vacuum fluctuations becomes
small and smaller. This situation may continue without any
nontrivial effect until a space-time ``continuum'' is formed out of
large enough numbers of discrete units of space-time so that a
``manifold'' can be effectively defined with an specific causal
structure which is called the ``metric''. In fact, since the
uncertainty principle is ignored for virtual quantum baby universes,
the causal deterministic structure can be formed out of these small
universes after their realization into the mother universe. It means
that at this stage of time evolution of the universe the rules of
standard ``quantum field theory'' govern on this new geometric
background. Suppose, by energy scale considerations, this stage of
time evolution of the universe coincides with the government of the
``electroweak'' interactions. The immediate question then arises
about the Higgs field which we believe gives mass to the standard
model particles. In this model, the only candidate for the Higgs
field at the phase transition from discrete to a continuous metric
structure of space-time is the ``scalar sector'' of vacuum quantum
gravitational fluctuations which determines the background energy
scale of the universe at that stage. In other words, once the
universe becomes large enough that the ``vacuum expectation value''
of a quantum field can be defined over the resultant space-time
manifold, the scalar sector of vacuum quantum gravitational
fluctuations with an energy scale much below the Planck scale,
namely the Higgs energy scale, may well effectively play the role of
a Higgs field. The fact that universal vacuum quantum gravitational
fluctuations may give mass to the particles is conceptually in
complete agreement with the Mach principle which is included in this
model\footnote{A similar idea is introduced in \cite{5'} where the
universal background Higgs field may be replaced by the background
gravitational potential of the universe.}. If so, there is no a real
Higgs particle because its associated field is just an emergent
background field out of quantum gravitational vacuum
fluctuations\footnote{The fact that mass of the Higgs particle is
not predicted by the standard model may be a confirming evidence
that it is not associated with a genuine field.}.

We have already assumed that expansion of the universe in space-time
is realized, step by step, by Planck length-time scales. We can
assume the same behavior for the mass extension of the universe.
Consider the following ratios
\begin{equation}
\frac{T}{t_P}\sim 10^{60}\:\:\: , \:\:\: \frac{R}{l_P}\sim
10^{60}\:\:\:, \:\:\: \frac{M}{M_P}\sim 10^{60} \label{2}
\end{equation}
where $T, R$, and $M$ are age, size, and observed mass of the
universe, respectively. We realize that in the Planck language the
amount of Planck masses in the universe is the same as the amount of
space-time units. This coincidence means that a one to one
correspondence exists between each Planck mass, each Planck length
and each Planck time. Therefor, in our Machian viewpoint of time
evolution of the universe during elapse of one Planck time the
radius of universe becomes larger with one Planck length and its
mass gets fatter with one Planck mass. There is no real problem with
this mass or energy generation because the universe is borrowing
this mass or energy from the vast gravitational vacuum sea to which
it is always contacted.

We may first study the generation of mass at early universe. Since
energy scale of the universe in contact with the vacuum
gravitational fluctuations is reducing by expansion of the universe
so the first mass generation had energy scale of the order of Planck
scale. Later, when the universe expands and its energy scale
decreases,  mass generation of ``reduced Planck'' scale happens. For
example, at the electroweak stage at which the energy scale of the
universe is about Higgs energy scale, the standard model particles
could be created\footnote{In the present model of mass generation
the massive particles are created first and the light particle
later. Therefore, the mass ratio of different particles depends on
the time difference (in the Planck unit $t_P$) between the
generation of those particles.}.

Note that at each stage of the history of the universe during one
Planck time the value of one Planck mass is generated. Let us, for
example, study the mass generation at the present stage of Machian
universe, namely with radius $R=10^{60} l_P$ and $T=10^{60} t_P$.
First of all, we have to find out the energy scale of the present
universe in contact with the gravitational vacuum. As explained
before, to evaluate the energy scale of the present mother universe
we have to use Heisenberg uncertainty principle. But, the very
important point is that since the universe as a four dimensional
object has both space and time extensions, we have to impose both
momentum-position and energy-time uncertainty relations at once on
the present universe and compare the result with that of imposed on
the original Planck length-time scale quantum baby universe, as
follows
\begin{equation}
(M_P^{reduced})^2 (R) (T) \simeq (M_P^{original})^2 (l_P)
(t_P),\label{*}
\end{equation}
where the units $\hbar=c=1$ are used. A simple calculation shows
$M_P^{reduced} \sim 10^{-60}M_P^{original}$ which implies that the
present energy scale of the universe is vanishing compared to its
initial value at the original stage of quantum baby universe. But
how this infinitesimal energy scale can lead to ``one'' Planck mass
generation at the present stage of Machian universe? Fortunately,
there are $10^{60}$ quantum baby universes with {\it temporal}
extension\footnote{Note that according to (\ref{2}) our large scale
universe has temporal and spatial extensions each of which $10^{60}$
times larger than those of each quantum baby universe. }which have
constructed the temporal extension of the present large scale
universe and each of them can contribute the same energy
$10^{-60}M_P^{original}$ into the present large universe. This gives
the desired ``one'' Planck mass $10^{60} \times
10^{-60}M_P^{original}=M_P^{original}$ generated per each Planck
time $t_P$. Note that by a Planck mass generation we do not mean a
Planck mass particle. In fact, generation of particles requires some
particle physics rules and symmetries other than their
space-time-mass characteristics. So, below the electroweak energy
scale toward the present very small energy scale we may not expect
the generation of stable standard model massive particles like
electrons and protons. However, neutrinos as particles with no
specific mass scale may be good candidates for the gradually mass
generation after electroweak stage for a while. This may justify the
present dark matter sector of the total mass $M$ of the universe. On
the other hand, the vast majority of the Planck mass generation may
be turned into the dark energy. In other words, after dark matter
generation such as neutrinos is completed, the vacuum energy of the
quantum gravitational fluctuations with ``reducing Planck'' scale
could be realized continuously into the expanding universe as the
dark energy.

The important results we obtained here are: 1) quantum gravity is
not a framework to quantize the gravitational interaction (force);
it is just a mechanism which continuously turns the ``vacuum quantum
gravitational fluctuations'' into an expanding universe, 2) quantum
gravity is fully Machian in that no space-time continuum exists in
the absence of gravitating matter; it creates space-time and matter
simultaneously, 3) each virtual single quantum baby universe after
realization into the mother universe no longer admits the
uncertainty principle, instead the mother universe does.

\section{Hierarchy problem}

It seems the energy scale of electroweak interaction is so below the
quantum gravitational interaction and the question that why this
happens is the mass hierarchy problem. In specific words, the
strength of a gravitational interaction between two particles
compared to their electroweak interaction is very small up to a
ratio $10^{-16}$. According to the present model for quantum gravity
one may justify such a large ratio by considering the very different
{\it space-time scales} of the two interactions and ask about the
hierarchy in their space-time scales. According to Heisenberg
uncertainty relation the energy scale of electroweak interactions is
$10^{3} Gev$. What about quantum gravitational interactions? We have
already reasoned that there is no such a real interaction and
quantum gravity just provides us with a space-time background so
that other quantum field interactions together with classical
general relativity can act on this background. So, how can we
measure the energy scale of gravitation to compare with electroweak
one? In fact, we do not need to measure the energy scale of a
quantum gravitational interaction between two particles, which does
not really exist. We just need to know the value of Newton's
gravitational constant which fortunately has already been fixed at
early universe by the original quantum baby universe as
$(\hbar=c=1)$
\begin{equation}
G=\frac{1}{M_P^2},\label{G}
\end{equation}
where $M_P \sim 10^{19} Gev$. So, it seems we are comparing the
parameters of ``electroweak universe'' with respect to that of the
original ``quantum baby universe'', as a reference. In other words,
we suppose the electroweak universe has happened right after the
appearance of quantum baby universe. The length-time scales for the
original quantum baby universe are $10^{-34}m , 10^{-43}s$,
respectively. On the other hand, we obtain the length-time scales
$10^{-18}m, 10^{-27}s$ for electroweak interaction. Now, during one
electroweak interaction in the ``electroweak universe'' the amount
of $\sim 10^{-18}/10^{-34} \times 10^{-27}/10^{-43} \sim 10^{32}$
space-time dimensional quantum baby universes contribute to the
original space-time according to the previously explained Mach
-Heisenberg principle. In fact, in order the electroweak interaction
can take place {\it during} $10^{-27}$ seconds over a {\it distance}
$10^{-18}m$, gravitational vacuum must deploy $10^{32}$ quantum baby
universes as the units of space-time each of which with the Planck
length-time scale $10^{-34}m-10^{-43}s$ and the reduced Planck mass
scale corresponding to the energy scale of that stage of the
universe. However, the energy scale of the universe at this stage is
assumed to be the electroweak energy scale. So, each quantum baby
universe created at this stage has energy scale $M_P^{reduced} \sim
10^{3} Gev$. It seems the gravitational $G$ looses its strength
against electroweak one because of the creation of a large amount
$\sim 10^{32}$ of space-time units to provide the length-time scale
that electroweak interaction needs to take place! To support this
idea we may provide the following relation
\begin{equation}
(M_{EW})^2 (L_{EW}) (T_{EW}) \simeq (M_P)^2 (l_P) (t_P),\label{EW}
\end{equation}
where Heisenberg's position-momentum and time-energy uncertainty
relations (in units of $\hbar=c=1$) have been used at once for the
electroweak universe in the left, and for the original quantum baby
universe, in the right. This relation leads to
\begin{equation}
\frac{G}{G_{EW}}\simeq \frac{M_{EW}^2}{M_P^2}\simeq \frac{(l_P)
(t_P)}{(L_{EW}) (T_{EW})}\simeq 10^{-32},\label{EW'}
\end{equation}
which shows the strength of gravitational coupling is $\sim
10^{-32}$ times weaker than electroweak one. Note that up to now we
have not really solved the hierarchy problem. We have just converted
the mass hierarchy to the space-time hierarchy. So at this point the
problem is why the space-time scales of electroweak universe is much
larger than the quantum baby universe? We have now a reasonable
justification for this hierarchy: we know the electroweak universe
is a stage in which massive particles are to be realized, namely the
energy scale of Higgs field is converted to massive particles. But,
no particle concept can be realized without the concept of
localization in space-time. So, we have to obtain a threshold
localization criteria for massive particles. In the present model,
the background vacuum gravitational fluctuations account for the
Higgs field giving mass to the particles. At the same time these
fluctuations are responsible for space-time realizations
corresponding to that energy scale, namely Higgs energy scale.
Therefore, the Higgs energy corresponding to a space-time scale can
be converted to a typical massive particle localized on this
space-time scale provided the Compton wavelength of the particle at
least is of the order of its size\footnote{If we consider the Higgs
mechanism as an effective model of this quantum gravity, this
criteria may play the role of order parameter which triggers the
electroweak spontaneous symmetry breaking so that the particles
become massive.}. The agreement of the Compton wavelength of the
proton with its actual radius is a clue to solve the space-time
hierarchy problem. It seems proton is the only stable standard model
particle which satisfies the localization criteria of a massive
particle. In other words, protons are the first massive particle
could be generated, since the birth of the first quantum baby
universe, by the background vacuum gravitational fluctuations,
namely the Higgs field\footnote{It is easy to show that a relation
like to (\ref{EW'}) exists to compare the Planck and proton energy
scales as
$$
\frac{M_{Pr}^2}{M_P^2}\simeq \frac{(l_P) (t_P)}{(L_{Pr})
(T_{Pr})}\simeq 10^{-39},
$$
where $M_{Pr}, L_{Pr}$ and $T_{pr}$ are the proton's mass, radius
and time light needs to pass the proton radius, respectively. The
large number $10^{-39}$ here has the same justification as explained
for the electroweak case. In fact, one can write down the
position-momentum and time-energy uncertainty relations in the
following times sequence for the quantum baby universe, electroweak
universe, proton universe and our large scale universe as
$$
(M_P)^2 (l_P) (t_P)\simeq (M_{EW})^2 (L_{EW}) (T_{EW}) \simeq
(M_{Pr})^2 (L_{Pr}) (T_{Pr}) \simeq (M_P^{reduced})^2 (R) (T).
$$
The last equality between the length-time-mass scales of a proton
and our present universe shows that proton is the unique particle
which can play the role of a small size universe. In other words,
proton universe is a typical small size universe which have led to
our present large scale universe. This sheds light on the puzzles
which Dirac had introduced in his large numbers hypothesis \cite{1}.
}. Before this criteria for massive particle production could be
realized, the universe just was enlarging filled with energies
pumped by vacuum gravitational fluctuations. Once this criteria is
realized the universe was then enlarging filled with massive
particles created by vacuum gravitational fluctuations. As the
universe expands according to Mach-Heisenberg principle, its energy
scale corresponding to the vacuum gravitational fluctuations,
decreases and so particles with less mass such as electrons are
realized whose Compton wavelengthes are larger than their actual
radiuses and so are more localized. In this way, all the standard
model particles could be realized during the electroweak era when
the electroweak universe was almost in contact with the vacuum
gravitational fluctuations with the energy scale of the Higgs
order\footnote{The fact that the actual masses of the fermions are
not predicted by Higgs-fermion couplings in the standard model is an
evidence that these couplings have their origin in the dynamical
coupling of the electroweak universe to the vacuum gravitational
fluctuations. In other words, the unknown couplings in the standard
model are related to the dynamical parameters of the evolving
universe in contact with the dynamical evolving vacuum energy scale.
The mass ratio of different particles will depend on the time
difference (in the Planck unit $t_P$) between the stages of
generation of these particles.}. After all the fermionic particles
with specific mass scales and bosons like photons as the $U(1)$
gauge particles are realized in the electroweak universe by
spontaneous symmetry breakdown, the universe is filled with matter
and radiation. The universe then continues its expansion getting
less energy scale from gravitational vacuum by Mach-Heisenberg
principle. At these later stages particles with variety of
decreasing mass or energy scales are to be realized whose best
candidates are neutrinos\footnote{In fact, these later and cooled
stages of the universe may coincide with matter-radiation decoupling
and cosmic microwave background CMB after which only light particles
such as neutrinos may survive leaving a background in neutrinos in
thermal equilibrium.} with unspecified mass scale which itself is a
good candidate for ``Dark matter''. Moreover, a large amount of
energy with decreasing scales can be gradually pumped (realized)
into this expanding universe by gravitational vacuum, which in turn
may be interpreted as what we call ``Dark energy''.

Now, we pay attention to the localization criteria and space-time
hierarchy problem. We may conclude that the reason why gravity
coupling $G$ is so weak against electroweak coupling $G_{EW}$ is
that electroweak era has to be realized in a stage of the universe
with such a large enough space-time scale on which the massive
particle can be localized over a space-time continuum. In other
words, the scale of causal and continuous structure of space-time at
electroweak universe over which a massive particle can be realized
should be very far enough (about $10^{32}$ order of scale) from the
scale of discrete structure of space-time at the stage of quantum
baby universe. This solves the space-time scales hierarchy and so
the mass scales hierarchy problems. This is like to the vast desert
between the continuous property of a light beam and its discrete
feature at the scale of one photon. For the same reason that one can
not realize a interference pattern by one or few photons, one can
not realize a particle concept by one or few quantum baby universes
(units of space-time). Rather, huge numbers of quantum baby
universes as eigenkets of a Hilbert subspace are required so that a
massive particle on a space-time continuum can be localized
according to the superposition rules of quantum mechanics.

We assert that in this model there is no time independent
unification of electroweak energy scale $M_{EW}\sim 10^{3} Gev$ and
Planck energy scale $M_P\sim 10^{19} Gev$. The Higgs sector is the
low energy limit of Planckian quantum gravity, namely quantum baby
universe, which is attained just by elapse of time or expansion of
the universe. In other words, at electroweak stage the energy scale
of decaying vacuum gravitational fluctuations with the reduced
Planck mass scale coincides with the Higgs energy scale
$M_P^{reduced} \sim M_{EW}\sim 10^{3} Gev$. Just in this regard, we
may have a unification.

\section{Cosmological constant problem}

If this model is capable of solving the hierarchy problem, it has to
be of the same capability to solve the cosmological constant
problem. The main prescription to solve the Hierarchy problem is to
consider carefully the space-time scales when we compare Higgs
energy scale with the Planck energy scale. The story is almost the
same when we wish to address the cosmological constant problem. Here
also we face with a problem which arises due to a comparison between
the current observed value of the vacuum (cosmological constant)
energy density and the quantum gravity contributions of the order of
$(M_{P})^4$. The ``4-th'' order huge correction $(M_{P})^4\sim
(10^{19})^4$ implies that something is wrong with the length-time
scales (in a ``4-d'' space-time) which have been used to evaluate
the cosmological constant today. So, we have to be very careful
about the length-time scales in this problem. According to above
prescription it seems these two energy scales are compared without
considering their characteristic length-time scales. In fact, in the
present era the characteristic length and time scales corresponding
to the present energy scale of the universe, namely the observed
cosmological constant, are the current size $\sim 10^{60}l_P$ and
current age $\sim 10^{60} t_P$ of the universe, respectively. The
characteristic length and time scales $l_P$ and $t_P$ do not belong
the the present stage of the universe rather they belong to the
early quantum baby universe. Fortunately, the Mach-Heisenberg
uncertainty principle lets us to relate and compare the
length-time-mass scales of one ``unique'' universe at these two
hierarchial stages as
\begin{equation}
(M_P^{reduced})^2 (R) (T) \simeq (M_P)^2 (l_P) (t_P),
\end{equation}
or
\begin{equation}
(\Lambda_c) (R) (T) \simeq (\Lambda_P) (l_P) (t_P),\label{3}
\end{equation}
where $\Lambda_c$ and $\Lambda_P$ account for the current observed
cosmological constant and the corresponding one at early quantum
baby universe, respectively. This relation asserts that each of the
cosmological constants $\Lambda_c$ and $\Lambda_P$ have their own
characteristic length-time scales belonging to two different stages
of the universe. So, if according to the popular interpretation of
quantum gravity we use scales $l_P$, $t_P$ and $M_{P}\sim 10^{19}
Gev$ to evaluate the quantum gravity contributions to $\Lambda_c$ we
will encounter with the well known cosmological constant problem,
because we are using the scales belonging merely to the early
universe and not present universe. If one is still interested in the
popular interpretation of quantum gravity to evaluate $\Lambda_c$,
he/she has to use the cosmological length-time scales $R$, $T$
leading to the reduced energy scale $M_P^{reduced}$. At the present
universe, the energy scale of quantum gravity (vacuum gravitational
fluctuations) is not the Planck scale $10^{19} Gev$, rather it is
very reduced due to the large expansion of the present universe.
Therefore, if one wishes to include the quantum gravity
contributions to $\Lambda_c$ at present universe he/she finds that
these contributions are of the same order of $\Lambda_c \sim
M_P^{reduced}$ and concludes that the current cosmological constant
is noting but the current ``background'' vacuum quantum
gravitational fluctuations to which our present universe is in
contact. This solves the cosmological constant problem.

Note that according to Mach-Heisenberg principle we assume that the
energy scale of each virtual single quantum baby universe realized
(socialized) in our present Machian universe is the same energy
scale of this present universe. So, one may compare the
length-time-mass scales of each virtual single quantum baby universe
at present and the one at early universes as ($c=1$)
\begin{equation}
t\simeq 10^{17} s  \left\{ \begin{array}{ll} l_P\simeq 10^{-34} m
 \\
t_P\simeq 10^{-43} s
\\
M_P^{reduced}\simeq 10^{-60} M_P,
\end{array}
\right.\label{present}
\end{equation}
and
\begin{equation}
t\simeq 10^{-43} s  \left\{ \begin{array}{ll} l_P\simeq 10^{-34} m
 \\
t_P\simeq 10^{-43} s
\\
M_P\simeq 10^{19} Gev.
\end{array}
\right.
\end{equation}
This latter certifies that Heisenberg uncertainty relation does not
work for a single quantum baby universe after realization
(socialization), according to Mach-Heisenberg principle, into the
mother universe. This is because as we probe the length-time scales
close to the Planck scales $t_P, l_P$ in the present universe we get
small energy scale $M_P^{reduced}$.

The story of cosmological constant problem is almost the same as Big
Bang problem: where did the Big Bang happen? The answer is:
``everywhere'' and ``nowhere''! If we take any point in the present
universe and trace back its history, it would start out at the
explosion point, and in that sense the Big Bang happened everywhere
in space. In another sense, the location of Big Bang is nowhere,
because space itself is evolving and expanding, and it has changed
since the Big Bang took place. Imagine the universe as an expanding
sphere. The place where the ``Bang'' happened is at the center of
the sphere, but that is no longer part of the space, the surface of
the sphere, in which we live \cite{6}. At present, we just observe
its imprint as ``CMB'' radiation with a very reduced temperature of
3.2 K. In the same way, the original cosmological constant of the
order of Plank mass squared has been very reduced due to the huge
expansion of the universe. So, if we wish to compute the quantum
gravitational corrections, namely the contributions of vacuum
quantum gravitational fluctuations to the cosmological constant, it
is so unreasonable to use the original Planck mass $10^{19} Gev$,
because this large mass scale has happened at the stage of quantum
baby universe and is no longer part of our present expanded
universe, so the Planck mass $10^{19} Gev$ is ``nowhere''. On the
other hand, its imprint at present universe is ``everywhere'' as a
very small mass scale $M_P^{reduced}\simeq (10)^{-60} M_P $ which we
interpret its square as the observed cosmological constant
$\Lambda_{reduced}\simeq (10)^{-120} \Lambda_P $.

\section{Discussion}

In this section we just speculate on the immediate impacts of this
model on the popular interpretation of quantum gravity and
cosmology. The result that a virtual single quantum baby universe
embed in our present mother universe do not admit Heisenberg
uncertainty relation is of particular importance in the study of
renormalizability of quantum gravity as well as other quantum field
theories. We know the quantum gravity is nonrenormalizable because
its coupling has dimension $(mass)^{-2}$ and the index of divergence
is growing by the order of perturbation. However, the important
point is that once we leave Heisenberg uncertainty relation for each
virtual quantum baby universe in our expanding mother universe we
realize that there are no large momentum-energy scales at ultrashort
Planck scales, rather we find that the energy or momentum scale is
infinitesimally small (\ref{present}). This means the degrees of
freedom which are relevant to the Planck scales are no longer
responsible for divergences. Therefore if one still insists to
interpret quantum gravity as an effective quantum field theory
he/she will find it free of ultraviolet divergences and a
renormalizable theory which may describe the creation of the
observed universe. For the same reason that uncertainty principle
looses it validity at small scales in an expanding universe, we have
no ultraviolet divergences corresponding to small scales in other
quantum field theories. In other words, it seems all the divergences
appearing in the standard quantum filed theories are due to the
assumption of a fixed mathematical space-time background in which
uncertainty principle is valid in this fixed background so that
large momentums appear at short scales. The renormalization technics
are then necessary to make these mathematical divergences finite.
However, in reality expansion of the universe as a natural
renormalization process makes all such quantum field theories to
lead to the finite results. In other words, in a physical expanding
space-time, rather than mathematical fixed space-time, in principle
we do not expect any real ultraviolet divergences.

The cosmology of this model is of Dirac-steady state type in which
mass or energy is continuously generated while the universe expands
in space and time homogeneously (see (\ref{2}) and the following
discussion) obeying {\it perfect cosmological principle} and {\it
large number hypothesis}. However, the present model has not the
problems of the original model which made it to be abandoned in the
literature.

The problem of continuous mass-energy generation is easily solved
because our universe is assumed to be in contact with the vacuum
state of quantum gravitational fluctuations and this vacuum sea
loans its energy in a continuous ``one-way'' to the universe, from
its original quantum baby state till its present large scale state,
according to the uncertainty principle which is always applied to
the whole Machian universe.

The problem of CMB is solved because in contrast to the original
model where all radiation in the universe originate in stars and
galaxies without thermalization to produce a perfect black body
spectrum, in this model all radiation (as well as matter) originate
from vacuum gravitational fluctuations which themselves are assumed
to be in a perfect thermalized state. Moreover, all radiation after
realization in the mother universe are in perfect causal contact
with each other because the rate of expansion of the universe in
this cosmology is always the light velocity $\frac{R}{T}=c$, so the
radius of universe always coincides with the horizon.

This model for quantum gravity, however, has major impacts on the
currently accepted scenarios for time evolution of the universe such
as inflation, Big Bang, and current acceleration of the universe. It
is important to note that the ratios $\frac{T}{t_P}\sim 10^{60} ,
 \frac{R}{l_P}\sim 10^{60}$ do not really mean a time evolution in
the present interpretation of the creation of the universe, because
space-time as a unique entity is generated dynamically. Rather, they
just tell us that the space-time extension of the present universe
is $10^{120}$ times larger than those of the original quantum baby
universe, that is it. These ratios do not tell us anything about
real time evolution. However, in this pure quantum gravitational
interpretation of the universe based on the fundamental constants
$c, G, \hbar$ we may interpret that the radius of the universe is
evolving with the constant light velocity $c$ in terms of a pure
quantum gravitational clock made of these constants as $t_P$. In
other words, we may consider this as the ``perfect uniform'' quantum
gravitational evolution of the universe with a fundamental time
parametrization made by quantum gravity itself, namely Planck time.
This is so reasonable because the vacuum sea of quantum
gravitational fluctuations which give birth and extension to this
universe is always in ``perfect thermalized'' state.

On the other hand, according to the present model there is a
fundamental distinction between quantum gravity and general
relativity. It turns out that quantum gravity is ``just''
responsible for the creation and perfectly uniform time evolution of
the Machian universe with all its stuffs inside, without telling us
how gravitational interactions take place between these stuffs
inside the universe. Here, Einstein equation plays the role and
``just'' tells us that how a massive object may curve the
surrounding space-time continuum and more importantly how this
curved space-time continuum determines the local geodesics. The
situation here is almost the same situation we have in the case of
Newton's laws of motion. We know the first law of Newton is ``just''
responsible for providing a well defined coordinate system in a
perfect uniform state of motion, namely inertial frame, so that the
second law of Newton becomes meaningful to ``just'' govern the
``local dynamics'' on this inertial background. In the same way, it
seems quantum gravity plays the role of first law of Newton to
provide a uniform Machian frame (space-time pavement) so that
Einstein equation (playing the role of second law of Newton) becomes
meaningful to describe the ``local dynamics'' on this Machian
background. Therefore, we arrive at the important conclusion that in
the framework of the present model, quantum gravity and Einstein
equation are different in essence \footnote{A same conclusion was
already proposed by the author elsewhere \cite{5}.}, almost in the
same manner that the first law of Newton is different from the
second law of Newton\footnote{The fact that just quantum gravity is
Machian sheds light on the useless effort of Einstein to reconcile
his local equation with Mach principle. In comparison, Einstein
would try to establish the first law of Newton, namely inertial
frames (Mach principle) by the ``vanishing force'' case of Newton's
second law (``vanishing energy momentum tensor'' case of Einstein's
equation).}. This distinction between quantum gravity and Einstein
equation provides us with opportunities to solve the remaining
problems in the present quantum gravitational interpretation of the
cosmology.

To this end, we again resort to the Newton's first and second laws.
If we take the rest inertial frame as our reference frame and wish
to describe the ``local dynamics'' of a particle with arbitrary
motion on this background we have to apply the second law of Newton
to obtain $\vec{r}(t)$. Now, suppose a particle is uniformly moving
with constant velocity $\vec{v}$ in the rest inertial frame as
$\vec{r}(t)=\vec{v}t$ and another observer in an arbitrarily
rotating frame with time dependent frequency and rotational
direction, observes the motion of this particle and wishes to
describe its local dynamics $\vec{r^{\prime}}(t)$. He/she will
certainly find a complicate motion for that particle in this
rotating reference frame and this is because the reference frame of
the observer has been changed from inertial to a rotating frame.
This complicate and nonuniform motion together with some terms with
wrong signs in the Hamiltonian to support the observations in this
rotating frame is the price the observer pays for the change of
reference frame.

We now identify the first and second laws of Newton with the quantum
gravity and Einstein equation, respectively. For the same reason
that one can not, in principle, use the second law of Newton to
alter the definition of ``uniform inertial frames'' in the first
law, here also one can not, in principle, use the local Einstein
equation to alter the quantum gravitational fundamental definition
of ``uniform evolution of the universe''. In other words, locally
one may use Einstein equation just to determine the local geodesics
on the surrounding space-time. On the other hand, for the same
reason that in the Newtonian dynamics the uniform motion of a
particle in an inertial frame may become a complicate motion if it
is observed in an arbitrarily rotating frame, the uniform expansion
(in time) of the universe (scale factor playing the role of the
particle) in the absolute Machian frame will appear as a nonuniform
(in time) expansion, if it is observed by a general relativistic
observer who uses Einstein equation to describe the ``global''
expansion of the universe. This is because the framework of general
relativity in comparison to that of quantum gravity is almost like
to the framework of a rotating frame in comparison to that of an
inertial frame.

To summarize, if we use Einstein equation (Newton's second law) in
the framework of general relativity (rotating frame) with a time
dependent ``equation of state'' (time dependent frequency) to
describe the evolution (motion) of the scale factor $R$ (particle)
having a uniform motion in the rest Machian frame (rest inertial
frame) defined by quantum gravity (Newton's first law), we obtain
complicate evolutions (motions) of the scale factor $R$ (particle),
such as inflation, Big Bang and current acceleration (acceleration,
deceleration and again acceleration), together with some terms in
the Hamiltonian with wrong signs in the kinetic term between
gravitational and matter degrees of freedom, and even potential
terms such as exotic matter or dark energy with ``negative''
pressure (frequency dependent terms with wrong signs)\footnote{Dark
energy is also predicted to be realized in our quantum gravitational
picture of the universe, but we do not need a negative pressure to
accelerate the universe because in this picture the universe is
uniformly expanding.} and violation of strong energy conditions (non
conservation of energy), to support the observations in the general
relativistic universe (rotating coordinate system)!

The point is that the initial ($R=l_P, T=t_P$) and final
($R=10^{60}l_p, T=10^{60}t_p$) states of the universe are the same
for both quantum gravitational (QG) and general relativistic (GR)
observers, but the ways and dynamics starting from initial state
toward ending at the final state are different. In other words, both
observers agree on a ``cosmological time'' which is set by quantum
gravity and in comparison to the Newtonian discussion made above we
may call it as the ``absolute time''. QG observer realizes a perfect
uniform evolution of the scale factor starting from initial and
ending at final state, according to this absolute clock. GR
observer, however, realizes accelerating (inflation)-decelerating
(Big Bang)-accelerating (current) phases for the same universe with
the same endpoints, according to the same absolute clock. This means
the appearance of nonuniform motions of the scale factor in the GR
frame is due to the nonuniform dynamic state of this frame with
respect to the QG rest absolute (vacuum) state frame. In fact, time
evolution of the GR universe depends on a dynamical state which is
set by different phases of its matter content, whereas time
evolution of the QG universe depends on an absolute state which is
set by perfectly thermalized vacuum quantum gravitational
fluctuations. In other words, the reason why GR observer experiences
nonuniform motions like inflation, Big Bang and current acceleration
in the GR universe is the existence of a ``dynamical state'' which
makes GR frame to be in nonuniform motion states with respect to the
QG frame with uniform motion and ``absolute state''. This behavior
for the GR universe is inevitable because unlike the QG absolute
gravitational universe, the GR universe is governed by new physics
coming from other three interactions which make its state of
matter-energy content to be dynamical which is usually determined by
the time dependent ``equation of state''\footnote{It is not so
difficult to construct, for example, a model universe whose dynamics
depends on the equation of state in four dimension while it has a
uniform constant dynamics independent of equation of state in higher
dimension \cite{5}. We may look for a similar pattern in the present
model to realize the difference between QG and GR universes if we
interpret the former and latter as 5 and 4 dimensional universes
(see the conclusion).}.

There is still more to learn. Suppose exactly in the same way that
local inertial forces manifest in a local rotating coordinate system
with respect to an ``absolute'' rest frame as Newton would demand,
universal inertial forces appear in the GR universe for a given GR
observer due to the nonuniform (in time) evolution of the GR
universe with respect to the ``absolute'' Machian universe. However,
since this nonuniform evolution of the scale factor is spatially
homogeneous and isotropic, the net effect of the universal inertial
forces imposed spherically from all directions on the observer is
zero and this observer is at rest. If this observer exerts a net
force in order to accelerate a particle at rest, the local balance
of universal inertial forces in the direction of acceleration will
change by this exerted local force and the nonuniform (accelerating
or decelerating) homogeneous and isotropic evolution of the universe
will try to fix the balance again. Therefore, the GR observer will
experience a restoring local force, imposed by nonuniform motion of
the universe while the particle is being accelerated, and will
interpret it as the ``inertia'' of the particle\footnote{It is
interesting that unlike Einstein original idea in considering an
static universe to recover the Mach principle, in this model
inertial properties of matter requires a universe with nonuniform
motion no matter this motion is accelerating or decelerating. The
rate of acceleration or deceleration does not change the
gravitational configurations. This is because we believe in the
Einstein equivalence between inertial and gravitational masses, so
all possible variations in the ``inertia'' of an object due to the
nonuniform (accelerating or decelerating) motion of the universe has
no detectable dynamical effect because of this equivalence. However,
different rates of acceleration or deceleration, for example, at
inflationary phase, Big Bang or current accelerating phases of the
universe leading to different values for the inertia of a particle,
may affect the non gravitational configurations and lead to
unexpected results.}. This is like the ``Mach force'' introduced for
the first time by E. Mach, but contrary to Mach's idea its origin is
not the gravitational effect of distant distribution of matter with
action of distance problem, rather it is the nonuniform homogeneous
and isotropic evolution of the universe which imposes this local
inertial force with no action of distance problem, because like the
local inertial forces which always appear simultaneously in the
local rotating systems the universal inertial forces, as well,
appear simultaneously throughout the entire universe as a coordinate
system with nonuniform motion.

Finally we may comment on two important puzzles in general
relativistic cosmology according to the present quantum
gravitational interpretation of the universe. First, since the
origins of matter and dark energy in the GR universe (according to
the present interpretation of the creation of universe based on
quantum gravity) are the same thermalized vacuum gravitational
fluctuations, this model then predicts that their order of
magnitudes should always be the same. It is easy to show that this
prediction is really the case. The matter density at present
universe is ($\hbar=c=1$)
$$
\rho_{_M}=\frac{M}{R^3}\sim 10^{-120}\frac{M_P}{l_P^3}\sim 10^{-120}
M_P^4
$$
and that of dark energy (cosmological constant) at present is
certainly the same
$$
\rho_{_{DE}}=\frac{\Lambda_c}{G}\sim
\frac{10^{-120}\Lambda_P}{M_P^{-2}}\sim 10^{-120} M_P^4.
$$
Therefore, the ``coincidence problem'' as an ``apparent'' problem in
the general relativistic interpretation of the universe is solved by
resorting to the quantum gravitational interpretation\footnote{In
fact, this is really a problem in GR interpretation while it is a
natural behavior in QG interpretation. By a change of reference
frame from GR to QG this problem is easily solved, like a complicate
and unjustified situation in a rotating frame which is easily solved
when we change our frame to an inertial one. }. Moreover, the fact
that the density of the total matter-energy content of the present
universe is divided up amongst the dark energy, dark matter and
baryonic matter as
$$
\rho_{_{DE}}>>\rho_{_{DM}}>\rho_{_{BM}},
$$
indicates that the ages corresponding to each content is set as
$$
\tau_{_{DE}}>>\tau_{_{DM}}>\tau_{_{BM}}.
$$
Therefore, the reason why the dark energy content has become
dominant to trigger the current acceleration in the general
relativistic interpretation of the universe is that according to the
quantum gravitational interpretation this energy is pumped
continuously into the universe so that at late time evolution of the
universe it is mostly filled with dark energy naturally, as
explained in section 4. Therefore the problem of ``dark energy
domination'' at the present general relativistic universe is solved
as well, by resorting again to the quantum gravitational
interpretation.
\newpage
\section*{Conclusion}

In this paper, we have proposed a new idea that space-time is
discrete and given a phenomenological model for quantum gravity
which we hope to solve the well known current problems of cosmology
and particle physics. In this interpretation, quantum gravity is not
an interaction, like the other three ones, mediating gravity force
at Planck scales. Rather, it is just responsible for continuously
generation of the universe (space-time-matter) by a newly developed
Mach-Heisenberg principle which lets virtual quantum baby universes
to be realized from a sea of vacuum quantum gravitational
fluctuations. The problem of time is solved due to the discrete
structure of space-time at ultra-short distances and a fundamental
time parametrization imposed by this quantum gravity. Hierarchy and
cosmological constant problems as the two problems with the same
origin are solved simultaneously by using Mach-Heisenberg principle
which makes the energy scale of the universe to decrease as it
expands, in this interpretation. It is shown that this quantum
gravity provides us with a framework in which all interactions
together with the vacuum quantum gravitational fluctuations may lead
to finite results without ultraviolet divergences, in an expanding
universe.

A very interesting prediction of this renormalizable quantum gravity
is that if two ultra high energetic particles collide each other up
to Planck energy scale they will be shattered into discrete units of
space-time, namely quantum baby universes of the Planck energy
scale. For simplicity suppose just one quantum baby universe is
generated in this collision. In fact, the Planck energy of this
quantum baby universe with Planck length-time scale is borrowed from
the energy content of our universe during the collision, but this
Planck energy per Planck time is provided in our universe by the
vacuum fluctuations. In other words, this time the universe borrows
one Planck energy per one Planck time from the vacuum, not for
itself (the whole universe) but for the generation of one quantum
baby universe with Planck energy. This new born quantum baby
universe, unlike the others already existed virtually in the mother
universe with reduced Planck mass, obeys naturally the uncertainty
principle firstly because it has been generated in a ``real
collision'' obeying uncertainty principle, and secondly because the
mother universe does not play the same game with this new quantum
baby universe. In fact, since this new quantum baby universe,
instead of generation by the vacuum, is generated inside the mother
universe, the mother universe has to support the Planck
length-time-energy scale of this quantum baby universe. In doing so,
the mother universe consumes of its own length-time-energy scale to
generate one Planck length-time-energy scale corresponding to the
new quantum baby universe. In other words, the mother universe does
not get old up to one Planck time, does not get large up to one
Planck length and does not get fat up to one Planck mass to
compensate for the generation of this quantum baby universe. In
simple terms, the mother universe stops its evolution for one Planck
time to compensate for the generation of this new quantum baby
universe. Therefore, this quantum baby universe seems to be rejected
by the mother universe back to the vacuum and so it plays the same
role of the original quantum baby universe which had led to the
present mother universe. In other words, this new born quantum baby
universe with huge energy scale is no longer part of our mother
universe with very small energy scale, rather it is rejected back to
the vacuum so as to be possibly the seed of another universe which
is to be expanded to another mother universe, causally disconnected
with us, by a new Big Bang!

We have shown that quantum gravity is completely distinct from
Einstein equation in that the former defines a uniform creation and
evolution of a Machian universe just like the first law of Newton
which defines just the rest inertial frames, whereas the latter
defines a nonuniform evolution of the general relativistic observed
universe just like the second law of Newton expressed in a
non-inertial time dependent frame, which describes the nonuniform
motion of a particle in this frame. It seems quantum gravity
establishes an ``absolute state of motion'' as Newton would desire
and general relativity establishes a ``Machian framework'' to
describe the universe as Einstein would desire.

The cosmological implications of this quantum gravity, based on
perfect cosmological principle and large number hypothesis, sheds
light on the most important subjects in cosmology such as Dirac
large number coincidences, time evolution of the universe like
inflation, Big Bang, current acceleration of the universe,
matter-energy content of the current universe and coincidence
problem. This model is free of singularity because a fundamental
quantum unit of space-time exists which never lets the scale factor
$R$ goes to zero. Moreover, horizon, flatness and fine-tuning
problems are easily solved in the quantum gravitational
interpretation of the universe. Since the quantum gravitational
universe is interpreted to be expanded always with the light
velocity we have not horizon and flatness problems. On the other
hand, there is no fine-tuning problem because there are no free
parameters in the quantum gravitational interpretation of the
universe except the fundamental constants $\hbar, c, G$. In fact,
the fine-tuning problem is specific to the general relativistic
universe with few parameters like density and deceleration
parameters. This model also gives a convincing justification for the
origin of ``inertia''.

Another interesting result is that the arrow of time in the observed
GR universe is induced by the arrow of time in the Machian QG
universe. Therefore, one may conclude that the reason why the
observed cosmological time in GR universe is not reversible is that
the arrow of time in the Machian QG universe is not reversible and
this is because the ``time'' is generated continuously through the
irreversible generation of quantum baby universes by Mach-Heisenber
principle which abandons the annihilation of these baby universes
after their realization into the causal Machian universe. However,
as explained in the second paragraph of the conclusion a
super-collision of the Planck energy scale may generate a few
quantum baby universes which may stop and reverse the direction of
cosmological time in our universe for a few Planck times. In other
words, this super-colliding local apparatus plays the role of a time
machine which may alter the cosmological arrow of time for a
while\footnote{Of course, it is not technically simple because in
order to stop and reverse the arrow of time for ``one second'' a
huge numbers of collisions of the order of $10^{43}$ is required!}.
Each of these missed Planck times in our universe may become a
beginning Planck time for other new universes!

If this phenomenological model works, one has to look for a
dynamical mechanism by which the quantum baby universes are
generated from vacuum gravitational fluctuations. A possible
mechanism would be generation of Planck scale wormholes occuring all
the time as virtual processes as conjectured by Baum, Coleman and
Hawking \cite{7}. On the other hand, since this model asserts that
the Higgs field as a genuine field does not really exist, it is of
particular importance to study a model by which the Higgs type
potential can effectively arise at ultrashort distances. One such a
mechanism is already proposed by the author \cite{5}. In this model,
an scalar field is defined at ultrashort distances (which
effectively may arise as the scalar sector of vacuum quantum
gravitational fluctuations) and a Higgs type potential is
effectively appeared. Non vanishing vacuum condensation of this
Higgs field occurs once a signature transition from Euclidean to
Lorentzian metrics is formed at ultrashort distances. It is
therefore appealing to suppose that quantum baby universes had
originally no demarcation between space and time. Once they have
been realized in a mother universe, their quantum uncertainty
features are faded in favor of a classical causal relation with each
other. We may interpret this as a signature transition from
Euclidean to the Lorentzian metric which could then lead to a large
effective vacuum condensation, in the effective Higgs type
potential, to trigger the inflationary era. This scenario is also in
the spirit of the ``no-boundary'' proposal of Hartle-Hawking in
quantum cosmology \cite{HH}.

It is worth noting that since this quantum gravity is a model of
space-time-matter generation, effectively it is in the spirit of 5D
{\it space-time-matter} theory developed by Wesson {\it et al}
\cite{Wesson}. The similarity may be realized roughly if one assumes
the vacuum gravitational fluctuations as a 5D entity, whose 4D
sector generates space-time and one remaining dimension as the
scalar sector generates matter (energy-mass), and assume this vacuum
state to play the role of 5D vacuum Ricci flat equation $R_{AB}=0$
which reduces to the Einstein equation with matter in 4D universe
almost like the realization of space-time-matter in our 4D universe,
according to the present model of quantum gravity. The emergence of
4D Einstein equation with matter from 5D vacuum Ricci flat equation
may imply that Einstein equation is emerged from Mach principle,
just like the emergence of Newton's second law from the first law.
Just in this manner we may have consistency between Einstein
equation and Mach principle. Note that, unlike brane higher
dimensional theories in which the 5th dimension plays an specific ad
hoc role to solve the cosmological or particle physics problems like
hierarchy or cosmological constant, the 5th dimension in this model
comes to play the natural role of matter. In this regard, we may
claim that our solutions to the above problems are natural.

\section*{Acknowledgment}

This work has been supported financially by Research Institute for
Astronomy and Astrophysics of Maragha (RIAAM).

\end{document}